\documentclass[twocolumn,superscriptaddress,amsmath,ammsymb]{revtex4-1}
\usepackage{epsfig}
\usepackage{amsmath}
\usepackage{amssymb}
\usepackage{graphicx}
\usepackage{dcolumn}
\usepackage{bm}
\usepackage[usenames]{color}
\usepackage[breaklinks]{hyperref}
\usepackage{soul}
\usepackage{ulem}
\hypersetup{colorlinks=true,allcolors=blue}

\begin{document}
	
\title{Magnetic field driven Lifshitz transition and one-dimensional Weyl nodes in three-dimensional pentatellurides}
\author{Zhigang Cai}
\affiliation{School of Science, Jiangnan University, Wuxi 214122, China}
\author{Yi-Xiang Wang}
\email{wangyixiang@jiangnan.edu.cn}
\affiliation{School of Science, Jiangnan University, Wuxi 214122, China}
\affiliation{School of Physics and Electronics, Hunan University, Changsha 410082, China}

\date{\today}

\begin{abstract} 
Recent experiments reported that the magnetic field can drive the Lifshitz transition and one-dimensional (1D) Weyl nodes in the quantum limit of three-dimensional pentatellurides, as they own low carrier densities and can achieve the extreme quantum limit at a low magnetic field.  
In this paper, we will investigate the conditions for the existence of the 1D Weyl nodes and their dc transport properties.  
We find that in the strong topological insulator (TI) phase of ZrTe$_5$, the formation of the Weyl nodes depends heavily on the carrier density; while in the weak TI phase of HfTe$_5$, the Weyl nodes are more likely to appear.  
These behaviors are attributed to the fact that in the strong and weak TI phases, the zeroth Landau levels exhibit opposite evolutions with the magnetic field.  
Moreover, the signatures of the critical fields that characterize the distinct behaviors of the system can be directly captured in the conductivities.  
\end{abstract} 

\maketitle

\section{Introduction}

When a perpendicular magnetic field is applied on a two-dimensional (2D) or three-dimensional (3D) electronic system, the electrons will be confined to move on curved orbits due to the Lorentz force~\cite{L.D.Landau}, and, consequently, a set of discrete energy levels, i.e., the Landau levels (LLs) will form.  If the magnetic field is strong and a few LLs are occupied, the system lies in the quantum Hall regime~\cite{K.V.Klitzing, D.C.Tsui, D.Yoshioka}.  
Further increasing the magnetic field, if all electrons occupy only the zeroth LLs, 
the extreme quantum limit will be reached. 
For most semiconductor materials, realizing the quantum limit seems impossible as the required magnetic field strength would be inaccessible in the experiment. 

In the past 20 years, the developments of topological materials have provided  insights to investigate the interactions between the magnetic field and electronic systems~\cite{M.Z.Hasan, X.L.Qi, Lv2021, J.A.Sobota}.  Among the emerging topological materials, 3D pentatellurides~\cite{Q.Li, T.Liang2018} including ZrTe$_5$ and HfTe$_5$ are representatives and can exhibit a number of desirable features: 
(i) the low-energy bands are well described by the effective noninteracting models that are topologically nontrivial~\cite{H.Weng}; 
(ii) they have narrow gaps and small band masses, leading to the sizable cyclotron frequencies even for a weak magnetic field~\cite{E.Martino, Z.G.Chen};  
and (iii) the crystal sample owns a high purity, with the electron mobility reaching the order of $10^5$ cm$^2$V$^{-1}$s$^{-1}$, and the low carrier density of about $10^{16}\sim10^{17}$ cm$^{-3}$ at a low temperature~\cite{F.Tang, E.Martino, P.Wang, S.Galeski2020}.  
Therefore, the quantum limit may be reached in 3D pentatellurides at a weak magnetic field, which makes its study possible.   

Recent experiments in pentatellurides reported that the system can indeed reach the quantum limit at the magnetic field of several Tesla~\cite{S.Galeski2022, W.Wu}.  
More importantly, there exists a magnetic field driven Lifshitz transition to the 1D Weyl regime: the crossing points of the zeroth LLs can be regarded as 1D Weyl nodes, because the band structure and spin texture are analogous to those of the Weyl nodes that are formed by Bloch band crossings~\cite{X.Wan, S.Y.Xu, Lv2015}.  
In this regime, since the Fermi surface is perfectly nested, the electronic states are unstable to the interactions~\cite{G.Gruner, F.Qin}, which makes it a good platform to investigate the strongly correlated electronic states.  
In ZrTe$_5$, the Lifshitz transition was characterized by the combined dc electric transport and ultrasound measurements~\cite{S.Galeski2022}.  Moreover, the chemical potential was found to meet and cross the Weyl nodes, resulting in the enhancement of the hole band occupation and the weakening of the electron band occupation~\cite{S.Galeski2022}. 
In HfTe$_5$, the Lifshitz transition was demonstrated by the magneto-infrared spectroscopy~\cite{W.Wu}, where a highly unusual reduction of optical activity and the variation of the accompanying resonant peaks were observed when the chemical potential crosses the zeroth LLs.  
Based on these observations, we propose the following questions: Do the Weyl nodes always exist in pentatellurides through modulating the magnetic field? If not, what are the conditions for their existence? 

In this paper, we will systematically explore the above questions in theory.  Although the ground states of ZrTe$_5$ and HfTe$_5$ depend heavily on the specific experimental conditions, such as the growth method, temperature~\cite{B.Xu}, and so on, it is widely believed that their ground states are located close to the phase boundary between a strong topological insulator (TI) and a weak TI~\cite{H.Weng}.  In the strong TI, the band inversions can occur in all three directions; whereas in the weak TI, the band inversion occurs only in one direction or in one plane.  More importantly, the two phases can be characterized by the Z$_2$ topological invariant index~\cite{L.Fu2007a,L.Fu2007b}.  Here we will take the ground state in ZrTe$_5$ and HfTe$_5$ as the strong and weak TI, respectively, which are supported by many experimental studies~\cite{Z.G.Chen, G.Manzoni, Y.Jiang, J.Wang, W.Wu}. 

Under the condition of fixed carrier density, we calculate the chemical potential as a function of the magnetic field $B$.  The chemical potential variation, together with the LL movements, will lead to several critical fields that characterize the distinct behaviors of the system.  We also study the dc transport property in the quantum limit by calculating the longitudinal conductivity $\sigma_{xx}$ and Hall conductivity $\sigma_{xy}$ by using the Kubo-Streda formula.  
We find that in the strong TI phase of ZrTe$_5$, the magnetic field can drive the zeroth LLs from crossing to be separated, thus the appearance of the 1D Weyl nodes depends heavily on the carrier density; 
whereas in the weak TI phase of HfTe$_5$, the magnetic field drives the zeroth LLs from separated to cross each other, which makes the 1D Weyl nodes more likely to appear.  
The critical fields can be captured by the peaks of $\sigma_{xx}$ and the vanishing $\sigma_{xy}$.  
Moreover, the linear dependence of $\sigma_{xy}$ on the inverse magnetic field, $\sigma_{xy}\sim B^{-1}$, holds in the quantum oscillation regime but does not in the quantum limit.  We attribute this observation to the $g_2$-spin Zeeman term that breaks the particle-hole symmetry. 
Our paper can help one better understand the quantum limit and the 1D Weyl nodes that are driven by the magnetic field in 3D pentatelluride experiments.

\section{Model and Methods}

We use the effective $\boldsymbol k\cdot\boldsymbol p$ model to describe the low-energy excitations in 3D pentatellurides.  In the four-component basis $\begin{pmatrix}
|+,\uparrow\rangle& 
|-,\uparrow\rangle& 
|+,\downarrow\rangle& 
|-,\downarrow\rangle
\end{pmatrix}^T$, the Hamiltonian is written as ($\hbar=1$) \cite{E.Martino, Y.Jiang, Z.Rukelj, J.Wang}
\begin{align}
H(\boldsymbol k)=&v(k_x\sigma_z\otimes\tau_x+k_yI\otimes \tau_y)+v_z k_z\sigma_x\otimes\tau_x
\nonumber
\\
&+[M-\xi(k_x^2+k_y^2)-\xi_zk_z^2]I\otimes \tau_z. 
\end{align}
Here $\sigma$ and $\tau$ are the Pauli matrices acting on the spin and orbit degrees of freedom, respectively.  $v$ and $v_z$ are the Fermi velocities, $\xi$ and $\xi_z$ are the band inversion parameters, and $M$ denotes the Dirac mass.  When taking $v_z=0$, $H(\boldsymbol k)$ becomes decoupled, and the upspin and downspin are good quantum numbers.  In the 3D system, there exist the inversion symmetry (IS) ${\cal I}^{-1}H(-\boldsymbol k){\cal I}=H(\boldsymbol k)$ with the operator ${\cal I}=\tau_z$, the time-reversal symmetry (TRS) ${\cal T}^{-1}H_0(-\boldsymbol k){\cal T}=H_0(\boldsymbol k)$ with ${\cal T}=i\sigma_y K$ and $K$ being the complex conjugation, the particle-hole symmetry (PHS) ${\cal P}^{-1}H(\boldsymbol k){\cal P}=-H(\boldsymbol k)$ with ${\cal P}=i\sigma_y\tau_x$, as well as the chiral symmetry (CS) ${\cal C}^{-1}H({\boldsymbol k}){\cal C}=-H(-{\boldsymbol k})$ with ${\cal C} =\sigma_y\otimes \tau_y$.  
Thus according to the Altland and Zirnbauer notations, the topological states described by $H(\boldsymbol k)$ belong to the chiral symplectic class CII~\cite{A.P.Schnyner, C.K.Chiu}. 

When a uniform magnetic field $\boldsymbol B=B \boldsymbol e_z$ acts on the 3D system, the 1D Landau bands will form, with the dispersions along the magnetic field direction.  To solve the LLs, we choose the vector potential in the Landau gauge as $\boldsymbol A=-By\boldsymbol e_x$ and make the Peierls substitution $\boldsymbol \pi= \boldsymbol k-e\boldsymbol A$.  Then we define the raising and lowering operators $a^\dagger=\frac{l_B}{\sqrt2}(\pi_x+i\pi_y)$ and $a=\frac{l_B}{\sqrt2}(\pi_x-i\pi_y)$, with $[a,a^\dagger]=1$ and the magnetic length $l_B=\frac{1}{\sqrt{eB}}=\frac{25.6}{\sqrt B}$ nm.  Besides the orbital effect, the magnetic field can also cause the spin Zeeman splitting, which is described as  
\begin{align}
H_Z=-\frac{1}{2}g_1\mu_BB\sigma_z-\frac{1}{2}g_2\mu_BB\sigma_z\tau_z.   
\end{align}
Here $\mu_B$ denotes the Bohr magneton, and $g_1$ and $g_2$ are the Land\'e $g-$factors.  
Since we have ${\cal P}^{-1}\sigma_z{\cal P}=-\sigma_z$ and ${\cal P}^{-1}\sigma_z\tau_z{\cal P}=\sigma_z\tau_z$, the PHS is preserved by the $g_1$ term but is broken by the $g_2$ term.    

With the trial wavefunction $\psi_n=(c_n^1|n\rangle, c_n^2|n-1\rangle, c_n^3|n-1\rangle, c_n^4|n\rangle)^T$, where the harmonic oscillator state $|n\rangle$ is defined by $a^\dagger a|n\rangle=n|n\rangle$ and $c_n^{1,\cdots,4}$ are the coefficients, we obtain the energies for the zeroth and $n\geq1$ LLs~\cite{L.You} 
\begin{align}
&\varepsilon_{0\lambda}(k_z)
=-\lambda\big(M-\xi_zk_z^2-\frac{\xi}{l_B^2}+\frac{1}{2}g_1\mu_BB\big)
+\frac{1}{2}g_2\mu_BB, 
\label{e0}
\\
&\varepsilon_{ns\lambda}(k_z) 
=s\Big[\big(M-\frac{2n\xi}{l_B^2}-\xi_zk_z^2-\frac{\lambda}{2}g_2\mu_BB\big)^2 +\frac{2nv^2}{l_B^2}\Big]^\frac{1}{2}
\nonumber\\
&\qquad\qquad\quad 
+\lambda\big(\frac{\xi}{l_B^2}-\frac{1}{2}g_1\mu_BB\big), 
\label{en}
\end{align}
respectively.  Here the index $s=\pm1$ denotes the conduction/valence band, and $\lambda=\pm1$ characterizes the upspin/downspin branch.  

With the density of states (DOS) $D(\varepsilon)$, the chemical potential $\mu$ is determined by the carrier density $n_0$ as~\cite{B.Fu, C.Wang}
\begin{align}
n_0=&\int_0^\infty d\varepsilon D(\varepsilon)f(\varepsilon-\mu)
+\int_{-\infty}^0 d\varepsilon D(\varepsilon)[f(\varepsilon-\mu)-1],  
\label{n0} 
\end{align}
where $f(x)=\frac{1}{\text{exp}(\beta x)+1}$ is the Fermi-Dirac distribution function with $\beta=\frac{1}{k_B T}$ the inverse temperature, and the charge neutrality is taken at the zero energy.

The contribution of the zeroth LL to the DOS is given analytically as 
\begin{align}
D_{0\lambda}(\varepsilon)
=\frac{g}{2\pi}\Big[\xi_z\big(C+\lambda\varepsilon-\frac{\lambda}{2}g_2\mu_BB \big)\Big]^{-\frac{1}{2}}, 
\end{align}
where $g=\frac{1}{2\pi l_B^2}$ is the LL degeneracy in the $x$-$y$ plane and can be denoted as the uniform DOS, $C=M-\frac{\xi}{l_B^2}+\frac{1}{2}g_1\mu_BB$.  We see that $D_{0\lambda}$ exhibits the square-root singularity, leading to the asymmetric peaks at $\varepsilon=-\lambda C+\frac{1}{2}g_2\mu_BB$.  

We further study the dc transport properties of the system, as the conductivities or resistivities can be measured directly in the experiments to help judge the electronic states.  The conductivity tensors are calculated by using the Kubo-Streda formula~\cite{L.Smrcka, G.D.Mahan}, 
\begin{align} 
\sigma_{\alpha\beta}=&\frac{1}{2\pi V}\sum_{\boldsymbol k} 
\int_{-\infty}^\infty d\varepsilon f(\varepsilon-\mu)
\Big[\text{Tr}\Big(J_\alpha\frac{dG^R}{d\varepsilon}J_\beta (G^A-G^R)
\nonumber\\
&-J_\alpha (G^A-G^R) J_\beta \frac{dG^A}{d\varepsilon}
\Big)\Big],    
\label{Kubo-Streda}
\end{align}
where $V$ is the volume of the 3D system, $J_\alpha=e\frac{\partial H}{\partial k_\alpha}$ is the current density operator along the $\alpha$ direction, and $G^{R/A}(\varepsilon,\eta)=(\varepsilon-H\pm i\eta)^{-1}$ is the retarded/advanced Green's function, with $\eta$ representing the LL linewidth broadening that is introduced phenomenologically to represent the impurity scatterings and will be taken as a constant for simplicity.  In the following, we focus on the zero temperature. 

With the help of the LL energies and wavefunctions, the longitudinal conductivity $\sigma_{xx}$ and Hall conductivity $\sigma_{xy}$ can be derived directly.  The selection rules $n\rightarrow n\pm1$ are determined from the nonvanishing matrix element of the current density, and there is no limit on the $s$ and $\lambda$ index.  We note that the conductivity components satisfy the following relations: 
\begin{align}
&\sigma_{xx}(ns\lambda\rightarrow n+1,s'\lambda')
=\sigma_{xx}(n+1,s'\lambda'\rightarrow ns\lambda),
\label{sigmaxx1}
\\
&\sigma_{xy}(ns\lambda\rightarrow n+1,s'\lambda')
=-\sigma_{xy}(n+1,s'\lambda'\rightarrow ns\lambda),
\label{sigmaxy1}
\end{align}
meaning that the contributions to $\sigma_{xx}$ from the LL transition $ns\lambda\rightarrow (n+1,s'\lambda')$ and from $(n+1,s'\lambda')\rightarrow ns\lambda$ are equal, while those to $\sigma_{xy}$ are opposite.  The expressions of $\sigma_{xx}$ and $\sigma_{xy}$ are obtained as~\cite{Y.X.Wang2023}
\begin{widetext}
\begin{align}
\sigma_{xx}=&\frac{\sigma_0\eta^2}{\pi^2 l_B^2} 
\int_{-\infty}^\infty dk_z \sum_{n\geq 0} \sum_{s,s'}\sum_{\lambda}
\frac{M_{ns,\lambda;n+1,s',\lambda}^2}{[(\mu-\varepsilon_{ns,\lambda})^2+\eta^2]
[(\mu-\varepsilon_{n+1,s',\lambda})^2+\eta^2]}, 
\label{sigmaxx2}
\\
\sigma_{xy}=&\frac{\sigma_0}{\pi l_B^2}\int_{-\infty}^\infty dk_z  
\sum_{n\geq 0}\sum_{s,s'}\sum_{\lambda}
\frac{[(\varepsilon_{ns,\lambda}-\varepsilon_{n+1,s',\lambda})^2-\eta^2] M_{ns,\lambda;n+1,s',\lambda}^2}
{[(\varepsilon_{ns,\lambda}-\varepsilon_{n+1,s',\lambda})^2+\eta^2]^2}
[\theta(\mu-\varepsilon_{ns}) 
\theta(\varepsilon_{n+1,s',\lambda}-\mu)
\nonumber\\
&-\theta(\mu-\varepsilon_{n+1,s',\lambda}) 
\theta(\varepsilon_{ns,\lambda}-\mu)],  
\label{sigmaxy2} 
\end{align}
where $\sigma_0=\frac{e^2}{2\pi}$ is the unit of the quantum conductivity, and $\theta(x)$ is the step function.  Explicitly, the matrix elements are given as 
\begin{align}
&M_{ns,1;n+1,s',1}=-\frac{\sqrt{2n}\xi}{l_B}c_{ns,1}^1c_{n+1,s',1}^1
+\frac{\sqrt{2(n+1)}\xi}{l_B} c_{ns,1}^2c_{n+1,s',1}^2+vc_{ns,1}^2 c_{n+1,s',1}^1, 
\label{M1}
\\
&M_{ns,-1;n+1,s',-1}=-\frac{\sqrt{2(n+1)}\xi}{l_B} c_{ns,-1}^3c_{n+1,s',-1}^3
+\frac{\sqrt{2n}\xi}{l_B}c_{ns,-1}^4c_{n+1,s',-1}^4-vc_{ns,-1}^3 c_{n+1,s',-1}^4. 
\label{M2} 
\end{align}
\end{widetext}
Due to the complicated form of the matrix elements $M_{ns,\lambda;n+1,s',\lambda}$, the integrations over $k_z$ in Eqs.~(\ref{sigmaxx2}) and~(\ref{sigmaxy2}) may not be completed analytically and need to be solved with numerics.

\section{Strong topological insulator in ZrTe$_5$}

\begin{figure*}
	\centering
	\includegraphics[width=17.2cm]{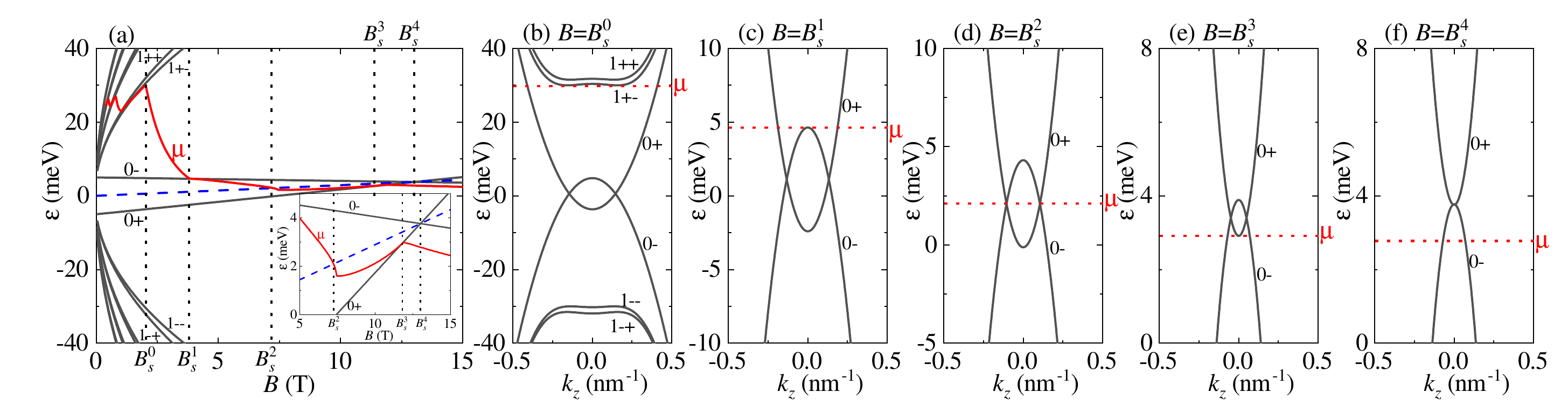}
	\caption{(Color online) (a) The LL spectra and the chemical potential $\mu$ in the strong TI, with the LL index $(ns\lambda)$ being labeled.  The carrier density is fixed at $n_0=6.76\times10^{16}$ cm$^{-3}$ and the critical fields are $B_s^0=2$ T, $B_s^1=3.85$ T, $B_s^2=7.25$ T, $B_s^3=11.75$ T and $B_s^4=13$ T, as indicated by the dotted lines.  The dashed blue line helps to judge $B_s^2$ and the inset shows the enlarged plot around zero energy. (b)$-$(f) The LL dispersions at the critical fields $B_s^{0,1,2,3,4}$, respectively, in which the dotted red lines denote the positions of $\mu$. }
	\label{Fig1}
\end{figure*} 

In this section, we study ZrTe$_5$ and take the model parameters from the experiment~\cite{Y.Jiang}: $M=5$ meV, $(v,v_z)=(6,0)\times10^5$ m/s, $(\xi,\xi_z)=(100,200)$ meV nm$^2$, $g_1=-8$ and $g_2=10$.  In fact, when the Fermi velocity $v_z$ is nonvanishing, a gap may be opened with the magnitude~\cite{Y.X.Wang2021} 
\begin{align} 
\Delta=2\sqrt{\frac{Mv_z^2}{\xi_z}-\frac{v_z^4}{4\xi_z^2}}. 
\end{align}
Here to observe the 1D Weyl nodes, we simply take $v_z=0$, so there is no gap opening and the system behaves as a semimetal.  Since the features of band inversions in the $x$-$y$ plane as well as the $z$ direction are kept and a finite $v_z$ will not affect the main conclusions of this paper (see Appendix A), we still regard the ground state of the system as a strong TI.  At a weak magnetic field, the band inversion in the $z$ direction leads to the zeroth LLs crossing at the momentum $k_z=\pm k_c$, with 
\begin{align} 
k_c=\Big(\frac{M}{\xi_z}-\frac{e\xi B}{\xi_z}+\frac{g_1\mu_BB}{2\xi_z}\Big)^\frac{1}{2}. 
\label{kc} 
\end{align} 

We fix the carrier density at $n_0=6.76\times10^{16}$ cm$^{-3}$ and calculate the chemical potential $\mu$ through solving Eq.~(\ref{n0}) self-consistently.  The results are displayed in Fig.~\ref{Fig1}(a) as a function of the magnetic field $B$, in which $\mu$ exhibits a non-monotonous variation.  In fact, $\mu$ is determined by the interplay between the local DOS due to the 1D LL dispersion and the uniform DOS $g$ that is proportional to $B$.  When $B$ is weak, $\mu$ lies above the $n\geq1$ LLs and the quantum oscillations are clearly visible in $\mu$.  According to the Onsager's relation, the oscillation period is closely related to the Fermi surface area~\cite{D.Shoenberg}, as  analyzed by us in a previous study~\cite{Y.X.Wang2023}.  When $B$ increases, the critical fields that are caused by the combined effects of the chemical potential variation and the LL movements will appear successively [Fig.~\ref{Fig1}(a)]. 

First, at the critical field $B_s^0$, the system enters into the quantum limit, with all electrons confined to the zeroth LLs [Fig.~\ref{Fig1}(b)].  When $B$ increases, the $0+$ and $0-$ LLs move upwards and downwards, respectively.  Since the uniform DOS gaining surpasses the local DOS dropping, $\mu$ decreases with $B$.  

At the critical field $B_s^1$, $\mu$ intersects the $0-$ LL [Fig.~\ref{Fig1}(c)], with 
\begin{align}
B_s^1=\frac{M-\mu}{e\xi-(g_1+g_2)\mu_B/2}. 
\end{align}
Then the Fermi surface varies from incorporating two points to four points.  Correspondingly, the crossing points at $k_z=\pm k_c$ own opposite chiralities and behave as two 1D Weyl nodes.  The effective Hamiltonian ${\cal H}^{\text{eff}}$ is written as 
\begin{align}
{\cal H}^{\text{eff}}=v_F(k_z\mp k_c)\sigma_z, 
\end{align}
where the Fermi velocity is 
\begin{align}
v_F=2\xi_z^\frac{1}{2} \Big(M-e\xi B+\frac{g_1\mu_BB}{2}\Big)^\frac{1}{2}.
\end{align}
Thus in the strong TI phase of ZrTe$_5$, the magnetic field drives the Lifshitz transition and  the system lies in the 1D Weyl regime.  

When $B$ further increases, $\mu$ decreases steadily, meaning that the magnetic field can effectively modulate the position of $\mu$ with respect to the Weyl nodes.  We see that $\mu$ will meet the Weyl nodes at the critical field $B_s^2$ [Fig.~\ref{Fig1}(d)], with 
\begin{align}
B_s^2=\frac{2\mu}{g_2\mu_B},     
\label{Bs2}
\end{align}
in which the $g_2$-spin Zeeman term plays a decisive role in driving $B_s^2$; without $g_2$, $\mu$ cannot meet the Weyl nodes.  In the experiment, $B_s^2$ is detectable only when $g_2$ is strong enough.  The decreasing $\mu$ is followed by an upturn [Fig.~\ref{Fig1}(a), inset].  This is because the $\Gamma$ point of the $0+$ LL moves to be above the zero energy, leading to the carriers turning from holes to electrons.  

\begin{figure}	
	\includegraphics[width=7.2cm]{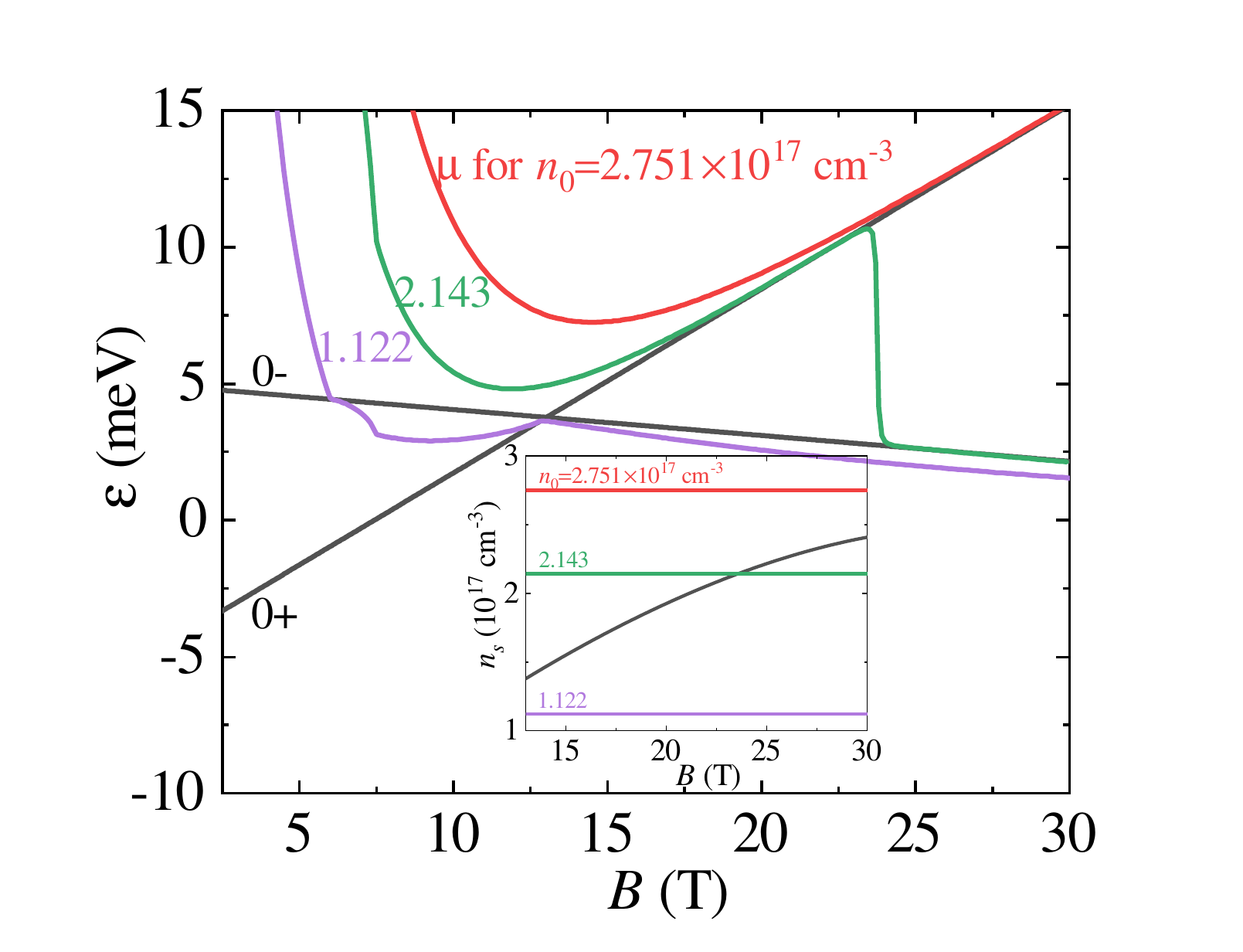}
	\caption{(Color online) The zeroth LLs in the quantum limit, and the chemical potential $\mu$ for a set of carrier densities $n_0=(2.751,2.143,1.122)\times10^{17}$ cm$^{-3}$.  The inset plots the characteristic carrier density $n_s$, with the horizontal lines denoting different $n_0$. }
	\label{Fig2}	
\end{figure}

\begin{figure*} 
	\centering	
	\includegraphics[width=17.2cm]{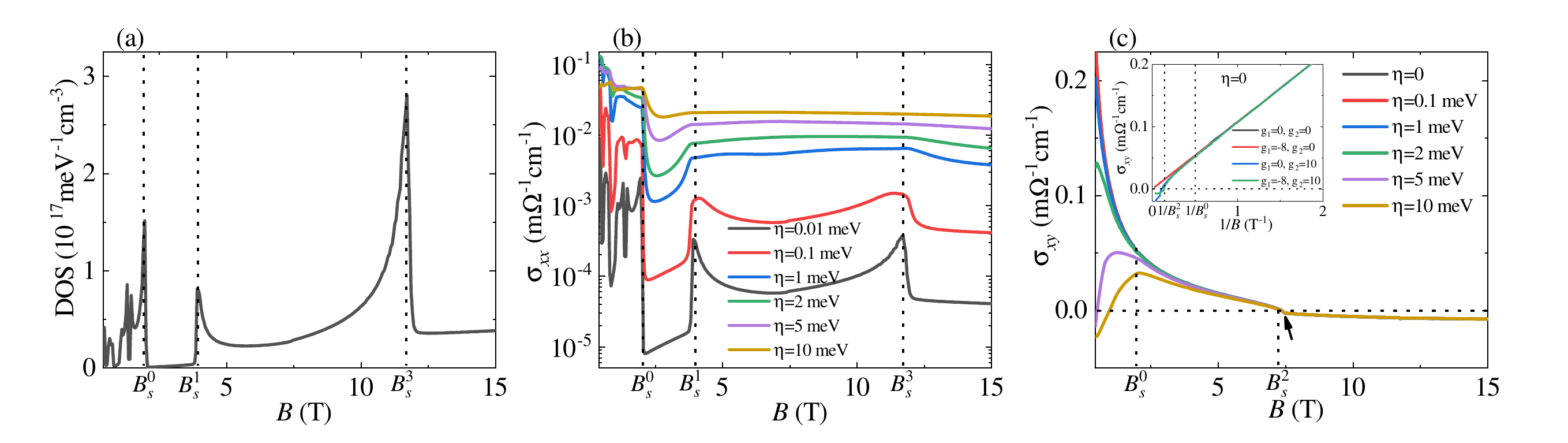}
	\caption{(Color online) The DOS (a), longitudinal conductivity $\sigma_{xx}$ (b) and Hall conductivity $\sigma_{xy}$ (c) versus the magnetic field $B$ in the strong TI, with the parameters the same as Fig.~\ref{Fig1}(a) and the critical fields labeled by the dotted lines.  (b) and (c) are plotted for different linewidths $\eta$.  The arrow in (c) marks the kink structure in $\sigma_{xy}$.  The inset in (c) plots $\sigma_{xy}$ versus the inverse magnetic field $B^{-1}$ at $\eta=0$, where the red dashed line shows the linear relation of $\sigma_{xy}\sim B^{-1}$, with the slope extracted as $k=0.108$ m$\Omega^{-1}$cm$^{-1}$T. }
	\label{Fig3}	
\end{figure*} 

After that, $\mu$ will be close to the $\Gamma$ point of the $0+$ LL and crosses it at the critical field $B_s^3$ [Fig.~\ref{Fig1}(e)], with 
\begin{align}
B_s^3=\frac{M+\mu}{e\xi-(g_1-g_2)\mu_B/2}.  
\end{align}
When $B>B_s^3$, only the $0-$ LL is occupied and the Weyl nodes do not exist.   

Finally, at the critical field $B_s^4$, with 
\begin{align}
B_s^4=\frac{M}{e\xi-g_1\mu_B/2}, 
\label{Bs4}
\end{align}
the zeroth LLs touch each other at the $\Gamma$ point [Fig.~\ref{Fig1}(f)].  Further increasing $B$, the zeroth LLs will be separated, and the system becomes a trivial insulator. 

According to the above analysis, in the strong TI phase, the magnetic field can drive the zeroth LLs from crossing to be separated.  Correspondingly, the system turns from topological nontrivial to trivial.  Based on this observation, we suggest that the 1D Weyl nodes do not always appear in the strong TI phase; their appearance depends heavily on the carrier density of the system.  To further clarify this point, we calculate the chemical potential $\mu$ for a set of the carrier density $n_0$ and plot the results in Fig.~\ref{Fig2}.  Since a gap is opened when $B>B_s^4$, if $\mu$ lies in the gap, only the $0-$ LL is occupied.  The characteristic carrier density $n_s$ is determined as 
\begin{align}
n_s(B)=\int_0^{\mu=\varepsilon_{0-}(\Gamma)} D_{0-}(\varepsilon) d\varepsilon
=\frac{g}{\pi}\sqrt{\frac{C+g_2\mu_BB/2}{\xi_z}}. 
\end{align}
The inset of Fig.~\ref{Fig2} shows that $n_s$ increases with $B$ monotonously. 

In Fig.~\ref{Fig2}, we see that there exist three cases for the 1D Weyl nodes:  
(i) When the carrier density $n_0<n_s(B_s^4)$, e.g., $n_0=1.122\times10^{17}$ cm$^{-3}$, the chemical potential $\mu$ intersects the $0-$ LL before the zeroth LLs getting separated, which enables the formation of the Weyl nodes.  This is similar to the case in Fig.~\ref{Fig1}(a).   
(ii) When $n_s(B_s^4)<n_0<n_s(B_m)$, e.g., $n_0=2.143\times10^{17}$ cm$^{-3}$, $\mu$ can only meet the $\Gamma$ point of the $0+$ LL after a gap is opened in the system.  Here $B_m$ represents the maximum magnetic field that would be accessible in the experiment and is taken as $B_m=30$ T.  Then $\mu$ will drop across the gap and lies below the $\Gamma$ point of the $0-$ LL, leading to the absence of the Weyl nodes.  
(iii) When $n_0>n_s(B_m)$, e.g., $n_0=2.751\times10^{17}$ cm$^{-3}$, $\mu$ cannot intersect the $0-$ LLs even after a gap is opened, and no Weyl nodes exist.  
Therefore, in the strong TI phase, the carrier density of the crystal sample directly determines the Weyl nodes that are driven by the magnetic field.   
For comparison, in Ref.~\cite{W.Wu}, a schematic plot of the chemical potential variation with $B$ is presented for a fixed carrier density in the strong TI phase, which is just case (iii).  

To find the signatures of the critical fields in the quantum limit, we calculate the DOS as well as the conductivity.  With the parameters chosen the same as Fig.~\ref{Fig1}(a), we plot the results in Fig.~\ref{Fig3} as functions of the magnetic field $B$, in which the critical fields are labeled by the dotted lines.   

In Fig.~\ref{Fig3}(a), in the quantum limit, three asymmetric peaks are exhibited at the critical fields $B_s^0$, $B_s^1$ and $B_s^3$, where $\mu$ crosses the $(1+-)$, $0-$, and $0+$ LL, respectively.  Explicitly, the peaks at $B_s^0$, $B_s^1$, and $B_s^3$ have long tails towards $B<B_s^0$, $B>B_s^1$, and $B<B_s^3$, respectively.  Moreover, the DOS remains finite when $\mu$ meets the Weyl nodes at $B_s^2$.  This is different from the Weyl nodes that are formed by breaking the TRS or IS in a Dirac semimetal~\cite{X.Wan, S.Y.Xu, Lv2015}, where the DOS vanishes at the Weyl nodes.  

\begin{figure*}
	\centering
	\includegraphics[width=17.2cm]{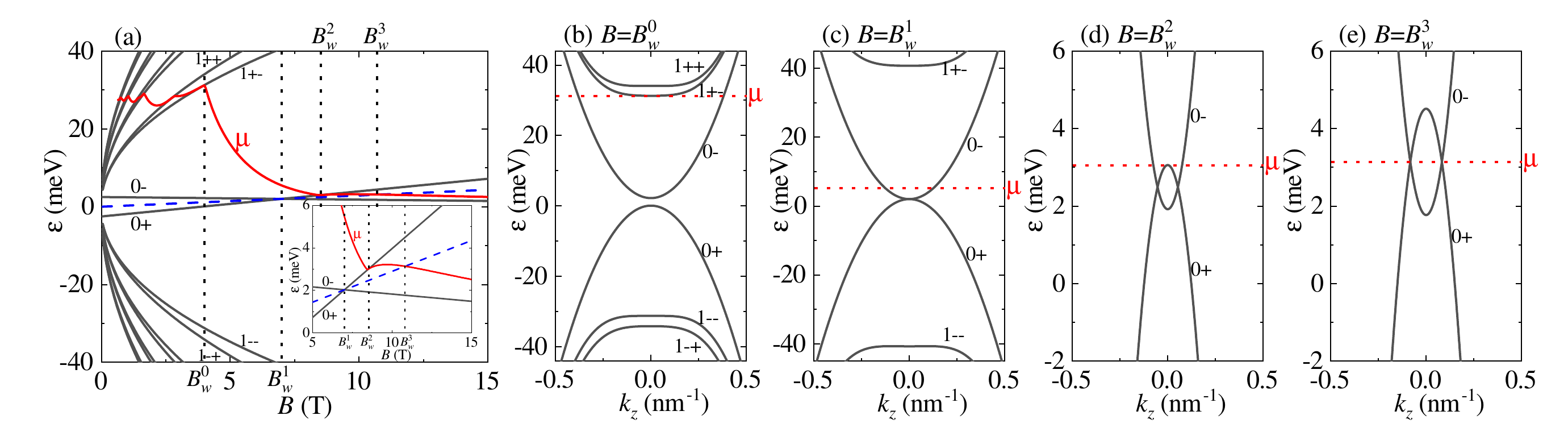}
	\caption{(Color online) (a) The LL spectra and the chemical potential $\mu$ in the weak TI, with the LL index $(ns\lambda)$ being labeled.  The carrier density is fixed at $n_0=1.25\times10^{17}$ cm$^{-3}$ and the critical fields are $B_w^0=4$ T, $B_w^1=7$ T, $B_w^2=8.6$ T, and $B_w^3=10.85$ T, as labeled by the dotted lines.  The dashed blue line helps to judge $B_w^3$, and the inset shows the enlarged plot around zero energy.  (b)$-$(e) The LL dispersions at the critical fields $B_w^{0,1,2,3}$, respectively, in which the red blue lines denote the positions of $\mu$. }
	\label{Fig4}
\end{figure*} 

In Figs.~\ref{Fig3}(b) and~\ref{Fig3}(c), the conductivities are shown, with the linewidth $\eta$ included to represent the effect of the impurity scatterings.  In the quantum limit, we observe that:  
(i) The longitudinal conductivity $\sigma_{xx}$ increases with $\eta$, whereas the Hall conductivity $\sigma_{xy}$ shows certain robustness to $\eta$.  These results are consistent with our previous studies~\cite{Y.X.Wang2020}.  
(ii) In $\sigma_{xx}$, corresponding to the DOS, three peaks are exhibited at the critical fields $B_s^0,B_s^1$ and $B_s^3$.  Such peaks are distinguishable when $\eta$ is weak but would be smeared at a strong $\eta=10$ meV. 
(iii) $\sigma_{xy}$ decreases smoothly with $B$ except a kink as marked by the arrow [Fig.~\ref{Fig3}(c)], which is attributed to the upturn behavior of $\mu$.  
(iv) $\sigma_{xy}$ vanishes at the critical field $B_s^2$ and will reverse its sign when the carriers change from electrons to holes.  The vanishing $\sigma_{xy}$ shows certain robustness to $\eta$, which favors the experimental observations.  Thus the signatures of $B_s^{0,1,3}$ and $B_s^2$ can be captured by the peaks of $\sigma_{xx}$ and the zero value of $\sigma_{xy}$, respectively.  But no signatures of $B_s^4$ are found in the conductivities, since the conductivities reflect the properties of the Fermi surface or Fermi sea in the system, and $\mu$ cannot meet the touching $\Gamma$ point of the zeroth LLs unless the carrier density is exactly $n_0=n_0^s(B_s^4)$.  In the experiment, the signatures of $B_s^1$ and $B_s^2$ have been reported in the dc transport measurements of ZrTe$_5$~\cite{S.Galeski2022}.  

In the inset of Fig.~\ref{Fig3}(c), we plot $\sigma_{xy}$ as a function of the inverse magnetic field $B^{-1}$ in the clean case $\eta=0$.  We see that in the quantum oscillation regime, the classical linear dependence of $\sigma_{xy}$ on $B^{-1}$~\cite{A.A.Abrikosov, V.Konye}, 
\begin{align}
\sigma_{xy}=\frac{n_0 e}{B},  
\end{align}
is retrieved. 
With the extracted slope $k=0.108$ m$\Omega^{-1}$ cm$^{-1}$ T, the carrier density is obtained as $n_0=6.75\times10^{16}$ cm$^{-3}$, which agrees well with the chosen value.  But in the quantum limit, $\sigma_{xy}$ shows evident deviations from the linear relation.  Intuitively, if the linear relation holds in the quantum limit, $\sigma_{xy}$ would vanish at an infinite magnetic field; now since $\sigma_{xy}$ vanishes at the critical field $B_s^2$, the linear relation can no longer hold. 

To further understand the above behaviors, we investigate the role of the PHS in $\sigma_{xy}$ by choosing a set of the Zeeman splittings $g_1$ and $g_2$ and plot the results in the inset of Fig.~\ref{Fig3}(c). 
When $g_1=g_2=0$, the system owns the PHS and the Weyl nodes are located at the zero energy.  We see that the linear dependence of $\sigma_{xy}$ on $B^{-1}$ holds in the quantum limit.  
When $g_1=-8$, the $0+$ and $0-$ LLs will move upwards and downwards, respectively, but keep  crossing each other.  As the PHS is preserved, the Weyl nodes remain located at the zero energy and the linear dependence holds in the quantum limit.  
However, at a finite $g_2=10$, the PHS is broken and thus the Weyl nodes are shifted in energy.  Consequently, the linear dependence will be destroyed in the quantum limit.  
In a recent magnetotransport experiment of ZrTe$_5$, the Hall resistivity $\rho_{xy}$ exhibited a similar dependence on the magnetic field~\cite{S.Galeski2021}, which supports our theoretical analysis. 

Actually, when the chemical potential $\mu>0$ lies between the $n$th and $(n+1)$th LLs, the dominant contributions to $\sigma_{xy}$ come from the LL transition $(n,1,\lambda)\rightarrow(n+1,1,\lambda)$.  This is reminiscent of the 2D Dirac fermion behavior in graphene~\cite{V.P.Gusynin}.  
When the PHS is broken, the $0+$ and $0-$ LLs will move asymmetrically in energy.  
On one hand, such movements will not change the topological property of the LLs, which can explain the vanishing $\sigma_{xy}$ at the Weyl nodes. 
On the other hand, in the quantum oscillation regime, $\sigma_{xy}$ is related to the transition of $n\geq1$ LLs, thus the LL movements will not affect $\sigma_{xy}$ and the linear relation; whereas in the quantum limit, $\sigma_{xy}$ is related to the transition of the zeroth LL to the $n=1$ LL, and the LL movements will change $\sigma_{xy}$ and destroy the linear relation.

\section{Weak topological insulator in HfTe$_5$} 

In this section, we study HfTe$_5$ and take the model parameters from the experiments~\cite{W.Wu}: $M=2.5$ meV, $(v,v_z)=(4.5,0)\times10^5$ m/s, $(\xi,\xi_z)=(120,-200)$ meV nm$^2$, $g_1=-6$ and $g_2=10$.  Now the ground state of the system is a weak TI that features the band inversion only in the $x-y$ plane.  Under a weak magnetic field, the topological trivial bands in the $z$ direction will lead to a gap between the zeroth LLs and the system behaves as a gapped insulator.  
At a fixed carrier density $n_0=1.25\times10^{17}$ cm$^{-3}$, the calculated chemical potential $\mu$ is displayed in Fig.~\ref{Fig4}.  The increasing magnetic field $B$ will drive the emergence of several critical fields [Fig.~\ref{Fig4}(a)]. 

\begin{figure*} 
	\centering	
	\includegraphics[width=17.2cm]{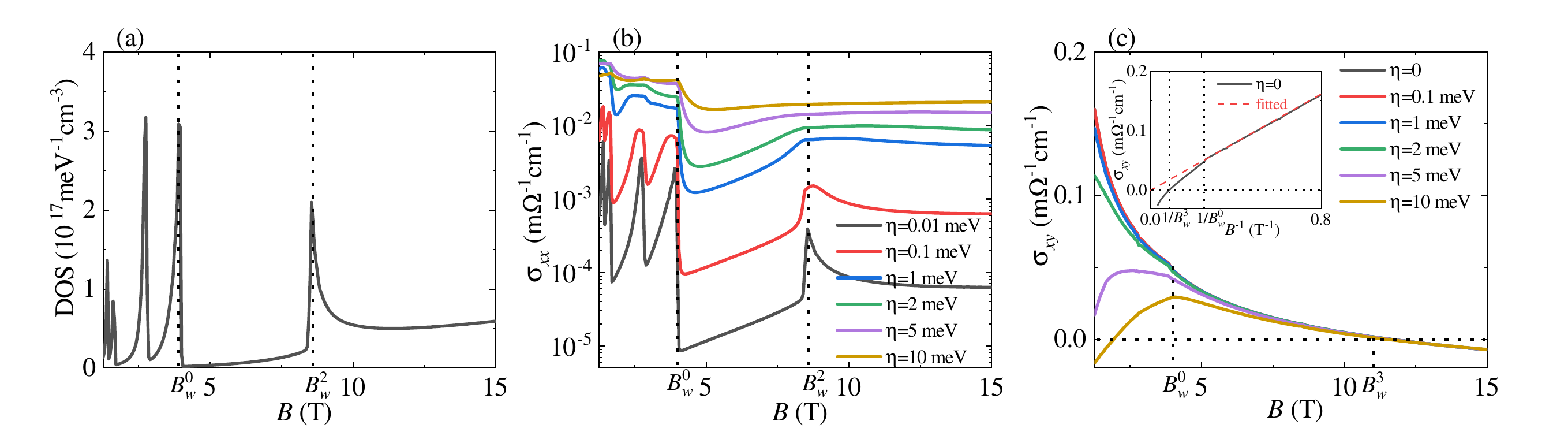}
	\caption{(Color online) The DOS (a), longitudinal conductivity $\sigma_{xx}$ (b) and Hall conductivity $\sigma_{xy}$ (c) versus the magnetic field $B$ in the weak TI, with the parameters the same as Fig.~\ref{Fig4}(a) and the critical fields labeled by the dotted lines.  (b) and (c) are plotted for different linewidths $\eta$.  The inset in (c) plots $\sigma_{xy}$ versus the inverse magnetic field $B^{-1}$ at $\eta=0$, where the red dashed line shows the linear relation of $\sigma_{xy}\sim B^{-1}$, with the slope extracted as $k=0.2055$ m$\Omega^{-1}$cm$^{-1}$T. }
	\label{Fig5}
\end{figure*}  

Firstly, the quantum limit is achieved at the critical field $B_w^0$ [Fig.~\ref{Fig4}(b)].  We see that the $0+/-$ LL lies in the valence/conduction band, which is different from the strong TI.  With increasing $B$, the $0+$ and $0-$ LLs will move upwards and downwards, respectively.  Since both the uniform DOS and local DOS increase with $B$, $\mu$ also decreases. 

At the critical field $B_w^1$, the zeroth LLs touch each other at the $\Gamma$ point so that the gap is closed [Fig.~\ref{Fig4}(c)].  Then the zeroth LLs cross each other at $k_z=\pm k_c$, with the gap being persistently closed.  Note that $k_c$ has the same expression as Eq.~(\ref{kc}).  When $B$ further increases, the decreasing $\mu$ will meet the $0+$ LL at the critical field $B_w^2$ [Fig.~\ref{Fig4}(d)].  
Meantime, the Lifshitz transition occurs in the weak TI phase, where the Fermi surface varies from incorporating two points to four points.  Correspondingly, the crossing points act as two 1D Weyl nodes. 
When $B>B_w^2$, $\mu$ shows an upturn and then decreases slowly.  This is because the local DOS drops when $\mu$ crosses the $\Gamma$ point of the $0+$ LL.  
Finally, $\mu$ will meet the Weyl nodes at the critical field $B_w^3$ [Fig.~\ref{Fig4}(e)].  

According to the above analysis, in the weak TI phase, the magnetic field drives the zeroth LLs from being gapped to cross each other.  Correspondingly, the system turns from topological trivial to nontrivial, which is opposite to the strong TI.  This is also seen from the fact that the critical fields $B_w^1$, $B_w^2$, and $B_w^3$ have the same expressions as $B_s^4$, $B_s^3$ and $B_s^2$, respectively.  Note that chemical potential variations with $B$ in the weak TI phase are consistent with Ref.~\cite{W.Wu}. 

It is interesting to ask whether there exists the critical field $B_w^4$ at which $\mu$ crosses the $0-$ LL.  When it happens, $B_w^4$ will have the same expression as $B_s^1$.  Now, since only the $0+$ LL is occupied, the characteristic carrier density $n_{w}^1$ is determined as   
\begin{align}
n_{w}^1(B)&=\int_0^{\mu=\varepsilon_{0-}(\Gamma)} D_{0+}(\varepsilon) d\varepsilon
\nonumber\\
&=\frac{g}{\pi} 
\Big(\sqrt{\frac{C-g_2\mu_BB/2}{\xi_z}}-\sqrt{\frac{2C}{\xi_z}}\Big),  
\label{n0m}
\end{align} 
which decreases with $B$.  Clearly, $B_w^4$ will appear if the carrier density satisfies $n_0<n_w^1(B)$ at a certain $B$.  For the magnetic field $B_w^1<B<B_m$, the calculations show that $n_w^1$ lies in the range $(0.96\sim5.45)\times10^{16}$ cm$^{-3}$, which is far below the chosen carrier density, thus $B_w^4$ would not appear. 

We investigate the conditions for the Weyl nodes in the weak TI phase. 
When the zeroth LLs get crossed, if the chemical potential $\mu$ meets the $\Gamma$ point of the $0+$ LL, the characteristic carrier density is 
\begin{align}
n_w^2(B)=&\int_0^{\mu=\varepsilon_{0+}(\Gamma)} 
[D_{0+}(\varepsilon)+D_{0-}(\varepsilon)] 
d\varepsilon
\nonumber\\
=&\frac{g}{\pi}\Big(\sqrt{\frac{C-g_2\mu_BB/2}{\xi_z}}+\sqrt{\frac{2C}{\xi_z}}\Big), 
\end{align}
which increases with the magnetic field.  For the magnetic field $B_w^1<B<B_m$, $n_w^2$ is given in the range $(5.45\sim133.87)\times10^{16}$ cm$^{-3}$, which is quite broad.  In the weak TI phase, the condition for the Weyl nodes is that the carrier density satisfies $n_w^1(B)<n_0<n_w^2(B)$ at a certain $B$.  Since this condition is easily satisfied, the Weyl nodes are more likely to appear in the weak TI phase.  

Next we study the DOS and conductivities of the system.  The results are displayed in Fig.~\ref{Fig5}, where the critical fields are labeled by the dotted lines.  At the critical fields $B_w^0$ and $B_w^2$, the chemical potential crosses the $(1+-)$ and $0+$ LL, respectively, which leads to the asymmetric peaks in the DOS as well as in the longitudinal conductivity $\sigma_{xx}$ [Figs.~\ref{Fig5}(a) and~\ref{Fig5}(b)].  The Hall conductivity $\sigma_{xy}$ decreases with $B$ smoothly and vanishes at the critical field $B_w^3$ [Fig.~\ref{Fig5}(c)].  Moreover, the effects of the linewidth $\eta$ on the conductivities are the same as those in Fig.~\ref{Fig3}.  Therefore, the signatures of the critical field $B_w^{0,2,3}$ are clearly seen in the conductivities.  In the inset of Fig.~\ref{Fig5}(c), we plot $\sigma_{xy}$ as a function of the inverse magnetic field $B^{-1}$ at $\eta=0$.  In the quantum oscillation regime, the linear dependence of $\sigma_{xy}$ on the inverse magnetic field $B^{-1}$ is seen.  With the extracted slope $k=0.1993$ m$\Omega^{-1} $cm$^{-1}$T, the carrier density is obtained as $n_0=\frac{k}{e}=1.244\times10^{17}$ cm$^{-3}$, which agrees well with the chosen carrier density.  But in the quantum limit, $\sigma_{xy}$ shows evident deviations from the linear dependence.  This observation can also be attributed to the broken PHS caused by the $g_2$-spin Zeeman term in the system, which is similar to the above strong TI phase analysis.  
In Ref.~\cite{W.Wu} of the longitudinal resistivity $R_{xx}$ measurements, a prominent linear behavior before saturating at high fields was observed but there were no evident peaks for the critical fields, thus more dc transport measurements in HfTe$_5$ are expected in the future.

\section{Discussions and Conclusions} 

\begin{figure} 	
	\includegraphics[width=7.5cm]{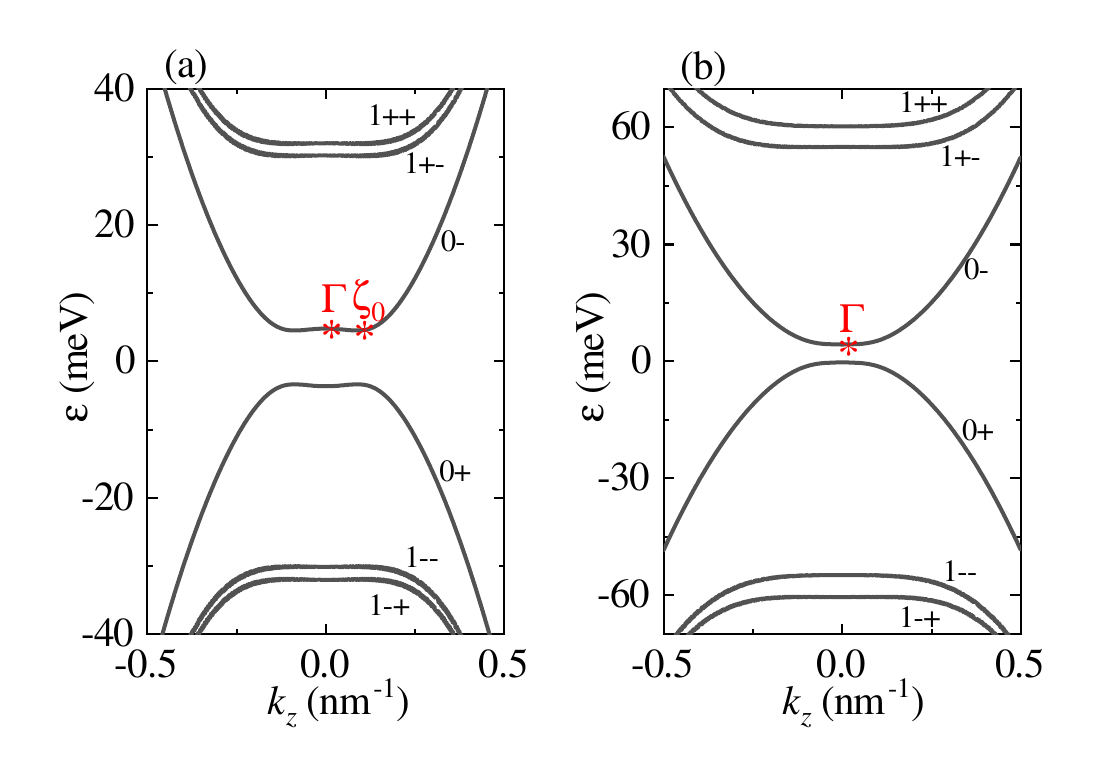}
	\caption{(Color online) The LL dispersions versus $k_z$ when the magnetic field $B=2$ T in (a) and $B=7$ T in (b), with the LL index $(ns\lambda)$ being labeled.  The Fermi velocity $v_z=5\times10^4$ m/s, and the other parameters are the same as Fig.~\ref{Fig1}(a).  The saddle points of the zeroth LL, $\Gamma$ and $\zeta_0$, are marked as asterisks. }
	\label{FigA1}
\end{figure}

In this paper, we focus on fixed carrier density $n_0$.  Now we discuss the case of fixed chemical potential $\mu_0$, since it can provide more insights to understand the exotic transport behavior in pentatellurides, such as the 3D quantum Hall effect~\cite{S.Galeski2021, F.Xiong, Y.X.Wang2023}.   
In both strong and weak TIs, with fixed $\mu_0$, the magnetic field can also drive the critical fields.  
In the strong TI, the Weyl nodes appear when $\mu_0$ satisfies  $\varepsilon_{0+}(\Gamma)<\mu_0<\varepsilon_{0-}(\Gamma)$, which strongly depends on the carrier density; 
while in the weak TI, the Weyl nodes appear when $\mu_0$ satisfies  $\varepsilon_{0-}(\Gamma)<\mu_0<\varepsilon_{0+}(\Gamma)$. 
The latter condition is easily satisfied, as the magnetic field drives the zeroth LLs to cross each other and the $0+$ LL move upwards.  Therefore we suggest that the conclusions for the 1D Weyl nodes with fixed $\mu_0$ are similar to those with fixed $n_0$.  

When the Fermi velocity $v_z$ is nonzero, the following consequences will be induced: 
(i) the 1D Weyl nodes will become gapped, thus the behavior of the zeroth LLs mimicks the physics of the 1D massive Weyl nodes; 
(ii) the upspin and downspin will be mixed and are no longer good quantum numbers;  
and (iii) in the strong TI phase, the zeroth LLs may avoid crossing each other, leading to the additional saddle points~\cite{Y.Jiang, J.Wang}.  Thus in the DOS and longitudinal conductivity $\sigma_{xx}$, more peaks will be found when the chemical potential crosses such saddle points.  The details are presented in Appendix A.  In addition, the effect of temperature on the magneotransport is briefly discussed in Appendix B. 

To summarize, our work explores the conditions for the magnetic field-driven Weyl nodes and the dc transport property in the quantum limit of the 3D pentatellurides.  Although the quantitative results depend on the model parameters, they are qualitatively valid and can show guiding significance for the experiments.  We hope that the 3D pentatellurides under a magnetic field can open an avenue for studying the interactions of 1D Weyl fermions as well as the resulting various strongly correlated electronic states.

\section{Acknowledgment}

This work was supported by the Natural Science Foundation of China (Grant No. 11804122), and the China Postdoctoral Science Foundation (Grant No. 2021M690970).

\section{Appendix} 

\subsection{LLs with nonzero $v_z$}

\begin{figure*} 
	\centering	
	\includegraphics[width=17.2cm]{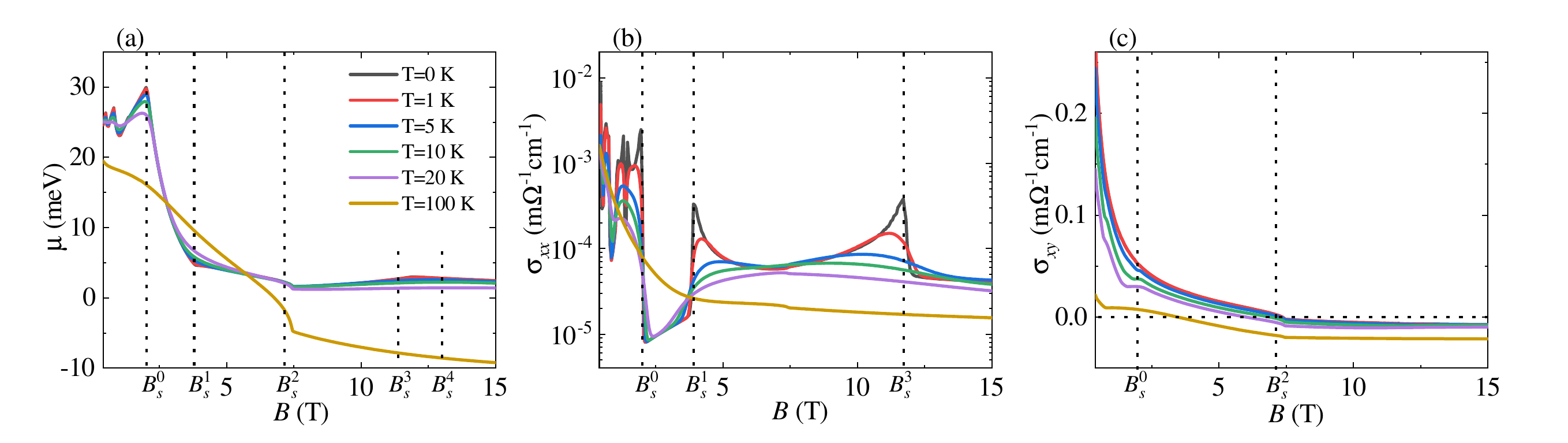}
	\caption{(Color online) The chemical potential $\mu$ (a), longitudinal conductivity $\sigma_{xx}$ (b) and Hall conductivity $\sigma_{xy}$ (c) versus the magnetic field $B$ for different temperatures $T$.  The parameters are the same as Fig.~\ref{Fig1}(a) and the critical fields are labeled by the dotted lines.  The legends are the same in all figures.  } 
	\label{FigA2}
\end{figure*}

For a finite Fermi velocity $v_z$, the zeroth LLs can still be obtained analytically, with the energies 
\begin{align}
\varepsilon_{0\lambda}(k_z)=&\lambda\Big[\big(M-\xi_zk_z^2-\frac{\xi}{l_B^2}
+\frac{1}{2}g_1\mu_BB\big)^2
+v_z^2k_z^2\Big]^\frac{1}{2}
\nonumber\\
&+\frac{1}{2}g_2\mu_BB, 
\tag{A1}  
\end{align}
while the $n\geq1$ LLs need to be solved numerically.  Now although the 1D Weyl nodes are gapped, the LL spectra at $k_z=0$ are unaffected by the finite $v_z$ in both strong and weak TIs [Figs.~\ref{Fig1}(a) and~\ref{Fig4}(a)], thus the evolution of the zeroth LLs with the magnetic field as well as the determined critical fields remain unchanged.  

In the strong TI, besides the saddle point $\Gamma$ at $k_z=0$, the additional saddle points $\zeta_n$ may appear~\cite{Y.Jiang, J.Wang, L.You}.  For example, when $B=2$ T [Fig.~\ref{FigA1}(a)], in the zeroth LLs,   the additional saddle points $\zeta_0$ can be found and are located at $k_z=\pm\big(\frac{M+g_1\mu_B/2}{\xi_z}-\frac{\xi}{\xi_zl_B^2}-\frac{v_z^2}{2\xi_z^2}\big)^\frac{1}{2}$.  With increasing $B$, $\zeta_0$ moves to $\Gamma$ and will finally merge with it at the critical field 
$B_{0c}=\frac{M-v_z^2/2\xi_z}{e\xi-g_1\mu_B/2}\simeq 6$ T.  When $B>B_{0c}$, there are no additional saddle points in the zeroth LLs [Fig.~\ref{FigA1}(b)].

\subsection{Effect of temperature}

Here we study the effect of temperature on the magnetotransport in pentatellurides.  At a finite temperature $T$, the conductivity $\sigma_{\alpha\beta}$ will be modified in two aspects: one is that $T$ can shift the chemical potential $\mu$; and another is that $T$ directly enters $\sigma_{\alpha\beta}$ via the Fermi-Dirac distribution function $f(x)$.  
By multiplying the zero-temperature conductivity $\sigma_{\alpha\beta}(T=0)$ by $\int_{-\infty}^\infty  d\epsilon\delta(\epsilon-\mu)$, the finite-temperature conductivity $\sigma_{\alpha\beta}(T)$ can be expressed as a weighted integration of $\sigma_{\alpha\beta}(T=0)$ around the chemical potential $\mu$ and is written as~\cite{L.Smrcka}
\begin{align}
\sigma_{\alpha\beta}(T)=\int_{-\infty}^\infty d\epsilon\sigma_{\alpha\beta}(T=0,\epsilon)
\big[-\frac{\partial f(\epsilon-\mu)}{\partial\epsilon}\big], 
\tag{A2}
\end{align}
where the derivative of the Fermi-Dirac distribution function is 
$-\frac{\partial f(\epsilon-\mu)}{\partial\epsilon} =\frac{1}{2k_BT\big(1+\text{cosh}\frac{\epsilon-\mu}{k_BT}\big)}$.

In Fig.~\ref{FigA2}, with the parameters the same as Fig.~\ref{Fig1}(a), we display the results of the chemical potential $\mu$, longitudinal conductivity $\sigma_{xx}$, and Hall conductivity $\sigma_{xy}$ for different temperatures $T$.  In Fig.~\ref{FigA2}(a), we see that with increasing $T$, when $B<B_s^0$ in the quantum oscillation regime, the oscillations of $\mu$ are weakened; when $B>B_s^0$ in the quantum limit, $\mu$ is less affected at $T\leq20$ K.  When the temperature increases to $T=100$ K, $\mu$ even becomes negative, meaning that $\mu$ shifts from the conduction band to the valence band.  This result is consistent with the experimental observations~\cite{Y.Zhang2017a, Y.Zhang2017b} and was believed to be the underlying physical mechanism of the anomalous resistivity peak at a finite $T$ in pentatellurides~\cite{C.Wang}.  In Fig.~\ref{FigA2}(b), the peaks of $\sigma_{xx}$ are smeared by temperature, and in Fig.~\ref{FigA2}(c), $\sigma_{xy}$ is gradually suppressed.  Note that when $T=100$ K, $\sigma_{xx}$ becomes a smooth curve,  indicating that the quantized LLs do not exist and the system enters the semiclassical diffusive region.

\end{document}